\documentstyle{lamuphys}
\makeatletter
\let\chapter\hid@chapter
\makeatother
\begin{document}
\pagenumbering{arabic}
\title{CP VIOLATION}

\author{Werner\,Bernreuther}
\institute{Institut f. Theoretische Physik, RWTH Aachen,\\
D-52056 Aachen, Germany\\ \\
Lectures given at the 37th Winter School on Particle Physics,\\
Schladming, Austria, Febr. 28 - March 7,  1998.}

\maketitle

\begin{abstract}
The salient features of CP-violating interactions in the 
standard electroweak theory and in a few of its popular extensions are discussed.
Moreover a brief overview is given on the status and prospects of searches 
for CP non-conservation effects in low and high energy 
experiments. 
\end{abstract}

\section{Introduction}
More than 30 years after the discovery of CP non-conservation in 
the neutral 
$K$ meson system neither the physics that causes this phenomenon 
has  been clarified nor have  other CP- or time-reversal-violating  effects been found
in laboratory experiments. The standard theory of 
electroweak interactions can explain the experimental findings by a complex
phase in the coupling matrix of the charged weak quark currents \cite{KM}. Yet in this theory  
the key to a deeper 
understanding of this symmetry violation is hidden behind the
mystery of the 
flavour problem. On the other hand, CP-violating interactions are 
conceivable which have nothing to do with the 
fact that there are three generations of quarks and leptons with
disparate mass spectra. Interactions of this kind 
naturally appear in popular and, so far, empirically
acceptable
extensions of the Standard Model. 
The question  whether
CP-symmetry breaking is due to a single 
``source" -- which is most likely the Kobayashi-Maskawa
phase \cite{KM} -- or whether there are several 
CP-nonconserving interactions which will show
up in different physical situations must be resolved experimentally. 
 Another enigma of particle physics --
often called ``problem number one" --  is the dynamics of 
electroweak symmetry breaking. Very probably
these two dark corners are related: clarification of weak gauge 
symmetry breaking -- which  must also be achieved by experiments --
 would also
shed light on the origin(s) of CP violation. 
\par In these lectures I shall first review the salient features
of CP violation in the Standard Model and in some of its 
extensions, notably multi-Higgs and supersymmetric extensions.
The issue of explicit versus spontaneous CP breaking will also be 
discussed. Then a brief overview is given
on the status and prospects of searches for CP non-conservation
effects in weak decays of strange, charmed, and beauty hadrons, on the search 
for permanent electric dipole moments
of particles, and on present and future high energy CP tests at colliders.

\section{Models}
The discussion in this section rests on the assumption that 
the Higgs mechanism -- which requires at least one 
 elementary scalar field multiplet with non-vanishing
ground state expectation value --
 gives the correct
description of electroweak gauge symmetry breaking. A priori, 
 breaking this symmetry does not require elementary
Higgs fields; it might have occurred ``dynamically" by condensation of (new) fermion bilinears. 
These
vacuum expectation values can have complex phases relative to each other, which induce  
observable
CP violation. 
I shall not 
discuss this 
possibility of CP-breaking, which is not without problematic features,
any further  here (see, e.g., \cite{Inagaki}).
\par
Moreover, the following discussion remains within the context of four-dimen\-sional 
gauge theories, where
CP constitutes a discrete symmetry transformation.
In higher dimensional theories, including string theories, CP can be
a gauge symmetry
which gets
spontaneously broken \cite{Dine}.
\par In the framework of four-dimensional gauge theories with elementary Higgs 
fields one can also distinguish between
two situations: \\
(a) CP invariance is violated explicitly at the Lagrangian level.
That is, in the ``Hamiltonian of the world", $H_{inv} + H'$, there is a term $H'$  (which by a posteriori
reasoning can usually be treated in perturbation theory) for
which $[H',U_{CP}] \neq 0$. Here
$U_{CP}$ is the unitary operator which implements the CP transformation
in the Hilbert space of the theory given by $H_{inv}$.\\
(b) One may have CP invariance of the full Hamiltonian, $[H,U_{CP}] = 0$, but
this symmetry is spontaneously broken by the ground state, $U_{CP} |0> \neq
e^{i\phi} |0>.$
This scenario requires more than one Higgs multiplet. In the following we shall discuss both options.

\subsection{The Kobayashi-Maskawa mechanism}
CP violation in the three-generation Standard Model (SM) of electroweak interactions
is related to the fact that  Nature has chosen the option that, as far as quarks
are concerned, the mass eigenstates are different from the
weak interaction eigenstates. (This may also be
the case in the lepton sector as recent experimental results on
atmospheric neutrinos and their interpretation in terms of
neutrino oscillations indicate.) In  the weak basis the Yukawa interactions 
of the quarks with the SU(2) doublet Higgs field are described by two  3$\times$3 coupling
matrices. The requirement of hermiticity of the Hamiltonian does not preclude that these matrices are
complex. After having transformed  the quark fields from the weak basis to the mass basis,
 the various pieces of the SM Lagrangian 
${\cal L}_{SM}$ are diagonal in generation space (in unitary gauge),
except for the  charged current interactions of quarks, 
\begin{equation}
{\cal L}_{cc} = - \frac{g}{\sqrt 2} \overline{U}_L \gamma^{\mu}V_{KM} D_L W^+_{\mu} +  {\rm h.c.} ,
\label{CC}
\end{equation}
which contains the Cabibbo-Kobayashi-Maskawa (KM) matrix \cite{KM},
a 3$\times$3 unitary matrix in generation space. Here ${\bar U} = ({\bar u},{\bar c},{\bar t})$ 
and $D = (d,s,b)^T$ are the quark
fields in the mass basis. 
Five parameters of the CKM matrix elements can be eliminated
 by a change of phase of the quark fields
\begin{equation}
u_i \to e^{i \omega_i} u_i ,\quad
d_j \to e^{i \tilde{\omega_j}} d_j \quad \Rightarrow \quad V_{ij} \to e^{i(\omega_i - \tilde{\omega_j})} V_{ij} ,
\label{phase}
\end{equation}
where i,j = 1,2,3 are generation indices. Hence 
the matrix $V_{KM}$ has four observable 
parameters, which may be chosen to be three Euler-type angles and 
a phase angle $\delta_{KM}$. If $\delta_{KM} \neq 0,\pm\pi$ the charged current 
Hamiltonian $H_{cc} = -\int d^3x {\cal L}_{cc}(x)$ is non-invariant under
a CP transformation.
\par
In view of eq. (\ref{phase}),  only 
functions of  $V_{ij}$ which are rephasing-invariant have a physical meaning. The simplest invariants
are $|V_{ij}|$ and 
\begin{equation}
Q_{ijkl} = V_{ij}V_{kl}V^*_{il}V^*_{kj}.
\label{Q}
\end{equation}
For three generations the unitarity of $V_{KM}$ implies that the $|{\rm Im}(Q_{ijkl})|$ are all equal \cite{Chau,Jarl}. 
(In fact the various unitarity triangles, representing the orthogonality relations of the
KM matrix elements, all have the same area which is equal to $|{\rm Im} Q|/2$.) Therefore, for instance
\begin{equation}
{\rm Im} Q = {\rm Im}(V_{ud}V_{cb}V^*_{ub}V^*_{cd})
\label{ImQ}
\end{equation}
is an invariant measure of CP violation \`a la KM. This expression immediately shows that
the strength of KM CP violation is small even if the CP-violating phase angle were maximal. 
Insertion of the 
measured values of the moduli of the KM matrix elements into eq. (\ref{ImQ}) yields that 
$|{\rm Im} Q|$ is smaller than $10^{-4}$. (For
a discussion of maximal CP violation in the SM, see \cite{Fritzsch}.)\\
A deeper understanding of CP violation \`a la KM requires an answer to the ``flavour problem", i.e.,
an answer to the question why there are three fermion generations and why is there such a hierarchy in 
the mass spectra of the u- and d-type quarks, respectively. If any two quarks with the same charge
were mass-degenerate, the CP-phase in $V_{KM}$ could be eliminated by a suitable unitary transformation of the
quark fields. This feature of KM CP violation is exhibited by the well-known invariant \cite{Jarl,Bra}
\begin{equation}
J_{CP} = \prod_{i>j \atop u,c,t} (m^2_i - m^2_j) \prod_{i>j \atop d,s,b} (m^2_i - m^2_j) 
\quad{\rm Im} Q .
\label{JJ}
\end{equation}
\par If neutrinos have non-degenerate masses then there can be KM-type CP violation 
in the lepton sector as well.
For three generations,
the leptonic analogue of the KM matrix, $V_{lept}$, which then parameterizes the relative strength of the
leptonic charged current-induced transitions, can have 1 CP phase angle (3 CP phase angles)
if the neutrinos are of the Dirac (Majorana) type.
\par
KM CP violation is observable only in flavour-changing charged current reactions. As is obvious from
eqs. (\ref{ImQ}) and (\ref{JJ}), effects are in general small because of  small mixing angles involved.
In charm hadron and in top quark decays, which are Cabibbo-allowed, even a maximal CP
phase in the KM matrix thus  leads only to very small effects. 
$K$ and $B$ mesons, whose weak decays are at least singly Cabibbo-forbidden, are therefore $the$ objects
to test the  KM mechanism.
\par A non-zero KM phase leads only to negligibly small effects in flavour-diagonal amplitudes.
For instance it induces  tiny electric dipole moments (EDM) of quarks \cite{Krause} and even tinier 
ones for charged leptons
(see section 4). 

\subsection{The strong CP problem}
At this point it is appropriate to recall the strong CP problem, which is actually
not a problem of Quantum Chromodynamics (QCD) in isolation, but of the  theory of strong $and$ weak interactions.
In QCD topologically non-trivial quantum fluctuations (instantons) induce a parity- and time-reversal-violating
term in the QCD  quantum action of the form 
$S_{\theta} = (\theta g^2/32\pi^2)\int d^4x G^a_{\mu\nu}{\tilde G^{a\mu\nu}}$, where $\theta$ is the
QCD vacuum angle. This term has  observable consequences in flavour-diagonal amplitudes.
Observables depend, however,  on the parameter
\begin{equation}
\overline{\theta} = \theta - arg(det{\cal M}_q)
\label{thetaQ}
\end{equation}  
where ${\cal M}_q$ is the non-diagonal mass matrix of the u- and d-type quarks in the weak basis.
The experimental upper bound on the neutron EDM implies \cite{Baluni} 
that $|\overline{\theta}| < 3 \times 10^{-10}$.
We lack a deeper understanding why this parameter should be so small.
Simply setting $\overline{\theta}$ equal to zero is unsatisfactory because, according to `t Hooft's 
naturalness condition \cite{Hooft},
it does not increase the symmetries of the SM. After all, CP must not be a good symmetry of the SM 
if this theory is to explain the observed CP effect
in $K^0$-$\overline{K^0}$ mixing. It requires $\delta_{KM}$ to be of order one 
and hence one would expect $arg(det{\cal M}_q)$ to
be of the same order. So there is apparently  severe fine tuning of $\theta$ required to 
bring $\overline{\theta}$ 
down to the level of $10^{-10}$.
For a more detailed discussion of this problem and of the possible ways out that have been proposed, 
see \cite{Peccei}. One may take a ``just so" attitude and consider $\overline{\theta}$ to be just another 
one of the
uncalculable parameters of the SM that happens to be (very close to) zero. 
However, many theorists believe that one cannot understand CP violation before one hasn't solved this problem.

\subsection{Extensions of the Standard Model}
There are a number of well-known arguments which motivate the belief 
in new physics beyond the Standard Model, to be discovered in particle
physics experiments. 
Extensions of the SM, even if based
on the gauge group $SU(3)_c\times SU(2)_L\times U(1)_Y$, almost invariably entail a 
larger non-gauge sector, that is to say, scalar self interactions 
and Yukawa interactions.
In this  way quite a number of ``new" CP-violating (CPV) interactions 
for quarks  $and$ for leptons
are conceivable in a natural way. 
(In the following, new CP-violating
interactions refer to interactions that are not due to the KM phase). 
In particular CPV interactions with the following features may exist:  \\
(a) Interactions that are unrelated to the mixing of quark generations and the hierarchy of quark masses. 
Such interactions induce CP effects also in flavour-diagonal amplitudes.\\
(b) Higgs-type interactions whose strength increases with the mass of the fermion involved, 
leading to sizeable effects in the heavy flavour sector.

\subsubsection{Explicit CP violation in multi-Higgs models}
The simplest, phenomenologically viable model with  
extra CPV besides the KM phase is, perhaps, the
extension of the SM by an extra $SU(2)$ Higgs doublet. 
The two Higgs doublets $\Phi_1, \Phi_2$ are assumed
to couple to quarks and leptons in such a way that there are 
no flavour-changing neutral couplings 
at the tree-level (see, e.g., \cite{Wein}). This
``natural flavour-conservation constraint" can be enforced
by imposing a discrete symmetry.
 The different implementations of this
symmetry define different 
models (see, for instance, \cite{Gunion}).
 Apart from complex Yukawa coupling matrices, which
lead to the KM phase, the requirement of hermiticity, renormalizability, and $SU(2)_L\times U(1)_Y$ 
invariance of
the  Lagrangian does not preclude explicit CPV in the Higgs 
potential $V_\Phi$. Requiring that the potential breaks the 
above-mentioned discrete symmetry only softly (that is, by terms with operator
dimension less than four) one can have
\begin{equation}
V_\Phi = V_0(\Phi_1,\Phi_2) + [\kappa \Phi^{\dagger}_1\cdot \Phi_2 + h (\Phi^{\dagger}_1\cdot \Phi_2)^2 +
{\rm h.c.}] .
\label{VH}
\end{equation}
Here $V_0$ denotes the CP-invariant part of the potential.  A CP transformation,
\begin{equation}
 \Phi_{1,2}(\mbox{\bf x},t) \quad \stackrel{CP}{\longrightarrow} 
\quad e^{i\alpha_{1,2}}\, \Phi_{1,2}^*(-\mbox{\bf x},t) ,
\label{VCP}
\end{equation}
shows that the term in the square brackets of eq. (\ref{VH}) breaks CP if $\xi ={\rm Im}(\kappa^2h^*) \neq 0$.
Note that it is unnatural to assume $\xi = 0$. Even if this were so at tree level, the
non-trivial KM phase $\delta_{KM}$, which is needed to explain the observed CPV, would induce a non-zero $\xi$
through radiative corrections. 
\par The spectrum of physical Higgs boson states in the two-doublet models consists of a charged Higgs
boson and its antiparticle, $H^\pm$, and three neutral states. As far as CPV is concerned, $H^\pm$ carries
the KM phase. It affects the CPV phenomenology of flavour-changing $|\Delta F| = 2$  neutral meson mixing
and $|\Delta F| = 1$ weak decays of mesons and baryons. From experimental data on $b \to s + \gamma$ the lower
bound $m_{H^+} > $ 210 GeV on the mass of this particle was derived \cite{Greub}.\\
If $\xi$ were zero, the set of neutral Higgs boson states would consist of two scalar (CP=1) and one pseudo-scalar
(CP= --1) state. Explicit CPV in the Higgs potential has the consequence that these states mix \cite{DeMa}
(note that
the mixing occurs already at tree level), leading to three mass eigenstates $\varphi_{1,2,3}$ that no longer have
a definite CP parity. That is, they couple both to scalar and to pseudo-scalar quark and lepton currents.
The Yukawa interactions read (for ease of notation the same symbol is used for a field and its 
associated particle state)
\begin{equation}
{\cal L}_{\varphi} = - (\sqrt 2 G_F)^{1/2} \sum_f (a_f m_f {\bar f}f + \tilde{a}_f m_f 
{\bar f}i\gamma_5 f) \, \varphi.
\label{Yphi}
\end{equation}
The sum over the Higgs fields $i=1,2,3$ is implicit.
Here $G_F$ is Fermi's constant, $f$ denotes a quark or lepton field, $m_f$ is the mass of the associated
particle, and the dimensionless reduced Yukawa couplings  $a_f, \tilde{a}_f$ depend on the parameters
of the Higgs potential and on the type of model \cite{BSP}. 
From LEP data one infers that the lightest of the three states $\varphi_i$ should have a mass larger than
about 50 GeV. (The precise lower bound depends on the parameters of the model.)
\par In terms of the parameters of eq. (\ref{Yphi}) CP violation in the neutral Higgs sector
occurs if $a_f\cdot\tilde{a}_f\neq 0$. The following generic features arise: (a)
The Yukawa interaction (\ref{Yphi})
leads to CPV in $flavour$-$diagonal$ amplitudes for quarks and for leptons. (b) The induced CP effects are 
proportional
to some power $(m_f)^p$, where one finds p=1,2,3 for reactions discussed in sections 4,5 below.
For example, neutral $\varphi$ exchange at tree level induces an effective CPV interaction
of the form $({\bar f}f)({\bar f}i\gamma_5f)$ with a coupling strength proportional
to $m_f^2$. Neutral $\varphi$ exchange at one-loop in the $\gamma ff$, $Zff$, and $G qq$ amplitudes 
($G$ denotes a gluon) leads to T- respectively CP-violating electric, weak, and chromo-electric dipole moment
form factors 
of the fermion involved which are proportional to $m_f^3$ \cite{BSP}. Potentially large effects can be expected 
for top quarks.
\par 
In models with a more complicated scalar sector, for instance, in models with $n\ge 3$ Higgs doublets, 
there is
more than one charged Higgs particle. The scalar potential can be such that these states mix
in a CP-violating way which leads to a complex mass matrix for these bosons.  Transforming all fields
to their respective  mass basis,
the interaction of the quarks with the charged Higgs bosons then reads  
\begin{equation}
{\cal L}_{ch} = - (2\sqrt 2 G_F)^{1/2} \sum_i (\alpha_i \bar{U}_L V_{KM} M_D D_R + \beta_i 
\bar{U}_R M_U V_{KM} D_L) \, H^+_i + {\rm h.c.},
\label{YHpl}
\end{equation}
where $M_{U,D}$ denote the real, diagonal 3$\times$3 mass matrices of the u- and d-type quarks,
and, in general,
\begin{equation}
{\rm Im}(\alpha_i\beta_i^*) \neq 0, 
\label{albet}
\end{equation}
due to the complex phases in the mass matrix of the charged Higgs bosons.
The interactions of
$H^\pm_i$ with leptons are of the same structure. 
In models where the right-handed quarks $q_R$ couple to several Higgs multiplets 
one can have additional CP phases such that the reduced Yukawa couplings
in eq. (\ref{YHpl}) satisfy
\begin{equation}
 \qquad {\rm Im}(\alpha_i\alpha_j^*) \neq 0,
\qquad {\rm Im}(\beta_i\beta_j^*) \neq 0 \quad (i\neq j) .
\label{ab}
\end{equation}
Charged Higgs exchange with couplings as in eq. (\ref{YHpl}) induces also CP violation in flavour-diagonal amplitudes.
For instance, if eq. (\ref{albet}) holds,
 one- and two-loop contributions to electric dipole form factors
of quarks and leptons  are induced (see section 4).  If eq. (\ref{ab}) holds,
  CPV chiral-invariant form factors in 
the $b{\bar b}ZG$ amplitude  are generated at one-loop order \cite{BBHN}.\\
In fact, even in two-Higgs doublet models  charged Higgs
boson exchange can provide CPV independent of the KM phase if
one allows for general Yukawa interactions \cite{WuL}
(which imply flavour-changing
neutral Higgs boson exchanges at tree-level).

\subsubsection{Explicit CP violation in supersymmetric models}
In the minimal supersymmetric extension (MSSM) of the Standard Model \cite{Nilles,HaKa} CP violating phases
can arise, apart from the complex Yukawa interactions of the quarks yielding a non-trivial
$\delta_{KM}$, from the soft  supersymmetry breaking terms.
The requirement of gauge invariance and hermiticity of the Lagrangian allows for \\
i) complex
Majorana masses $M_i$ in the gaugino mass terms, 
\begin{equation}
-\sum_i (M_i\lambda_i\lambda_i)/2 + {\rm h.c.}, 
\label{gaug}
\end{equation}
ii) complex trilinear scalar couplings, 
that is, complex 3$\times$3 matrices ${A}_{q,\ell}$ in generation space
which contain the couplings
of the
scalar quarks and leptons to the  Higgs doublets $\Phi_1,\Phi_2$. 
For instance\footnote{In order to facilitate the comparison
with the above models, the non-SUSY
convention, i.e., the same hypercharge assignment for both  $SU(2)$ Higgs doublets,  $\Phi_i = (\phi^+_i, \phi^0_i)^T$, (i=1,2)
is employed here.} 

\begin{equation}
 {\tilde{D}}'^\dagger_R{A}_d \Phi^\dagger_1\cdot{\tilde{Q}}'_L + {\rm h.c.},
\label{tri}
\end{equation}
and analogous interactions with coupling 
matrices ${A}_{u}$ and ${A}_{\ell}$.
The tilde and the prime 
denote scalar quark
fields in the weak basis, capital letters denote as before  vectors in generation
space. The label L
refers to $SU(2)_L$ doublets,  ${\tilde{Q}}'_L = ({\tilde{U}}'_L, 
{\tilde{D}}'_L)^T$, and the
label $R$ in eq. (\ref{tri}) refers to $SU(2)_L$ singlets. \\
iii) a complex soft term in the Higgs potential
\begin{equation}
 \kappa \Phi^\dagger_1\cdot\Phi_2 + {\rm h.c.} 
\label{bterm}
\end{equation}
Motivated by supergravity models it is often assumed that the $A_f$'s are proportional to the
Yukawa coupling matrices
\begin{equation}
{A}_f = A{Y}_f, \qquad f = u,d,\ell ,
\label{Aff}
\end{equation}
where $A$ is a complex parameter. The observable CP phases of the MSSM are readily counted \cite{Dug}. Apart from the KM phase
and the QCD $\bar{\theta}$ parameter, these are $arg(AM^*_i)$ and $arg(\kappa M^*_i)$, where $M_i$ are the gaugino
mass terms in eq. (\ref{gaug}). If eq. (\ref{Aff}) is not
imposed then there are quite a number of independent phases.
\par 
After spontaneous symmetry breaking and after having transformed the various fields such that all 
mass matrices have become real and diagonal, the CP phases
have been shifted into the fermion-sfermion-neutralino
and -chargino interaction terms. Let us write down here only the gluino interaction, which
involves the QCD coupling $g_{QCD}$.
One arrives at the CP-violating quark-squark-gluino interaction Lagrangian in the mass basis,
which reads for u-type quarks
\begin{equation}
{\cal L}_{{\tilde G}u\tilde u} = i{\sqrt 2}g_{QCD}
\sum_{j,l} (e^{-i\varphi_u}{\bar u_{jL}}\Gamma_{jl}{\tilde G^a}
T^a {\tilde u_l} + e^{+i\varphi_u}{\bar u_{jR}}\Gamma_{jl}'{\tilde G^a}
T^a {\tilde u_l}) + {\rm h.c.} ,
\label{gluino}
\end{equation}
where j=1,2,3, and l=1,...,6, $\tilde G^a$ denote the gluino fields,
 $T^a$ are the generators of $SU(3)_c$
in the fundamental representation, and $\Gamma,\Gamma'$ are complex 3$\times$6 matrices.
(Recall that for each flavour there are two squark respectively slepton mass eigenstates, which are in
general not mass-degenerate.) The ${\tilde d}d\tilde G$ interaction is of the same form.
As already mentioned there are further  CP-violating fermion-sfermion-neutralino and -chargino
interactions (of similar structure as eq. 
(\ref{gluino})) with interesting phenomenological consequences.
\par
If eq. (\ref{Aff}) holds then the phase $\varphi_q = \varphi_{\tilde G} - \varphi_A$ is universal and $\Gamma,
\Gamma'$ depend, as far as CP phases are concerned, only on the KM phase. 
However, in the general case things are really more complex. \\
As the gluino interactions involve both flavour-diagonal and flavour-changing $\Delta Q = 0$ vertices,
${\cal L}_{{\tilde G}q\tilde q}$ induces CPV effects in neutral meson mixing, in flavour-changing decays
of hadrons, and it leads to electric dipole moments (EDM), e.g. of the neutron, of considerable size. The latter
constitutes a well-known conflict for the MSSM. The predictions for the neutron EDM come out too large as
compared with the experimental upper bound
 if the CP phases of the soft terms i) ii) and iii) above
were of order one and if the  squark and gluino masses were about, say,  200 GeV. (See section 4).
Therefore it is often assumed in the literature that 
the CP phases of the soft terms i) ii) and iii)
are  zero at a very high energy scale, which is usually taken 
to be  a supposed  grand unification scale or the Planck scale.
  Then CP violation at this scale is assumed to come  only from the Yukawa couplings, i.e., the KM phase. When evolving
the parameters of this  constrained MSSM model down to lower energies, the KM phase induces, through
renormalization, also small phases in the soft SUSY breaking terms \cite{Berto}, which are phenomenologically
acceptable as far as EDMs are concerned.
(For a discussion of the  CP-violating phases and their phenomenological consequences in the supersymmetric grand 
unified $SO(10)$
model, see \cite{Hall}.)
\par
What about Higgs sector CPV in supersymmetric extensions of the SM? In the MSSM 
the tree-level Higgs potential is, schematically,  of the form 
\begin{equation}
V_{tree} = V_0(\Phi_1,\Phi_2) + (\kappa \Phi^{\dagger}_1\cdot \Phi_2  +
{\rm h.c.}) .
\label{Vsusy}
\end{equation}
As is well-known (see, e.g. \cite{Gunion}) SUSY does not allow for
independent  quartic couplings in $V_0$. They 
are proportional to linear combinations of the $SU(2)$ and $U(1)$ gauge couplings squared. Moreover,
 a term 
of the form $(\Phi^{\dagger}_1\cdot \Phi_2)^2$ and two other quartic terms which are non-invariant under
$\Phi_1 \to - \Phi_1$  are absent at tree-level. Hence by suitable adjustment of the
phases $\alpha_i$ in eq. (\ref{VCP}) a CP transformation on the scalar fields can be implemented such that
$\int d^3x V_{tree}$ is CP-invariant. Thus there is no CPV mixing of the three physical neutral Higgs states
at tree level. The CPV interactions in the MSSM discussed above generate a 
(small) complex coupling $h$ (cf. eq. \ref{VH})  in the effective potential 
at one-loop order, and the parameter $\xi$ defined above can now become non-zero. The other quartic terms absent at
tree-level will also be induced. Hence there can be 
CPV mixing at one-loop order of the two CP=1 and the CP= --1 neutral Higgs states in the MSSM.
 It is, however, 
expected to be small.
\par
In  next-to-minimal SUSY models with two $SU(2)$ Higgs doublet fields, $\Phi_1, \Phi_2$, and a gauge
singlet field $N$ the Higgs potential  can explicitly violate CP at the tree level. For instance, this is
the case for the potential
\begin{equation}
V_{tree} = V_{inv}(\Phi_1,\Phi_2,N) + (h_1N^3 + h_2\Phi^{\dagger}_1\cdot \Phi_2N +h_3\Phi^{\dagger}_1\cdot \Phi_2N^2 +
{\rm h.c.}) ,
\label{VNSSM}
\end{equation}
where $h_{1,2,3}$ are arbitrary complex couplings and $V_{inv}$ is the CP-invariant part of 
the potential. Appropriate field redefinitions show that there is  one observable
CPV phase in eq. (\ref{VNSSM}). In this case 
the three CP=1 and the two CP= --1 physical neutral Higgs states mix at tree-level \cite{Mats}.
(There is, however, no mixing of the two scalar with the pseudo-scalar component of the two
Higgs doublets.)

\subsubsection{Spontaneous CP violation}

There is a potential cosmological problem when  discrete symmetries 
 are spontaneously broken
\cite{Zeld}. When spontaneous CPV occurs in the early universe at some temperature, one expects 
that domains with different
CP signatures (i.e., different signs of the CP-violating phase(s)) are formed. The energy density of the  walls which
separate these domains dissipate not rapidly enough when the universe expands further. Estimates yield that
 the energy density associated with these walls today exceeds the closure density of the universe by many orders of 
magnitude \cite{Zeld}.  The problem is avoided if CP got broken before inflation took place. However, 
the connection to low energy 
phenomena then becomes very loose.

Ignoring this problem, it is nevertheless
interesting to investigate multi-Higgs or supersymmetric extensions of the SM with spontaneous
CPV at the weak scale. 
The simplest model in this respect is the original two-Higgs doublet model of T.D. Lee \cite{Lee}, 
respectively its more recent variants \cite{Bran,Wolf,WuL}.
By construction the Lagrangians of these models, which have  the same gauge group and the same
particle content --
apart from the Higgs sector -- as the SM, are CP-invariant.
 Hence the Yukawa couplings can be taken to be real, without loss of generality. 
Gauge invariance, hermiticity, and renormalizability imply that the tree level Higgs
potential has the general form \cite{Lee}
\begin{eqnarray}
V \, = \, V_0(\Phi_1,\Phi_2) 
+ (\kappa \Phi^{\dagger}_1\cdot \Phi_2 + \lambda_1 (\Phi^{\dagger}_1\cdot \Phi_2)^2 +
\lambda_2 (\Phi^{\dagger}_1\cdot \Phi_2)(\Phi^{\dagger}_1\cdot \Phi_1) \nonumber \\
\quad + \lambda_3 (\Phi^{\dagger}_1\cdot \Phi_2)(\Phi^{\dagger}_2\cdot \Phi_2)
+ {\rm h.c.}) , 
\label{VLee}
\end{eqnarray}
where, contrary to eq. (\ref{VH}), the parameters in eq. (\ref{VLee}) are real because $V$ is required to be
CP-invariant. \\
Minimisation of the  potential yields the vacuum expectation 
values (VEV) of the two Higgs fields (the phase of $\Phi_1$ has been adjusted such that
the VEV of this field is real)
\begin{equation}
<0|\phi^0_1|0> = v_1/\sqrt 2, \qquad <0|\phi^0_2|0> = v_2 e^{i\vartheta}/\sqrt 2, 
\label{VEV}
\end{equation}
where $v_1, v_2$ are real and positive parameters which have to satisfy the experimental constraint
$\sqrt{v^2_1 + v^2_2} = $ 246 GeV. 
There exists a range of parameters  
of $V$ such that the absolute minimum is characterized by a 
non-trivial phase \cite{Lee}
\begin{equation}
\vartheta = \arccos[\frac{2\kappa - \lambda_2 v_1^2 - \lambda_3 v^2_2}{4 \lambda_1 v_1 v_2}].
\label{theta}
\end{equation}
The necessary condition for this to happen is
\begin{equation}
\lambda_1 > 0 , \qquad |\frac{2\kappa - \lambda_2 v_1^2 - \lambda_3 v^2_2}{4 \lambda_1 v_1 v_2}| < 1 .
\label{cond}
\end{equation}
If  the phase angle $\vartheta \neq \pm n \pi/2$, n = 0,1,2,.., then
the ground state breaks CP spontaneously. It can be shown that
there is then no unitary implementation 
of CP such that the Lagrangian and the vacuum remain invariant \cite{BrGG}.
\par One consequence of spontaneous CPV is, again, CPV mixing of neutral Higgs states. 
Yet one must account for the observed CPV in $ K^0$-${\bar K^0}$ mixing, but the Yukawa couplings are real by 
construction. Therefore the construction principle of ``natural flavour conservation" must
be given up.
The right-handed quark fields $u_{iR}, d_{iR}$ are coupled
both to $\Phi_1$ 
and $\Phi_2$,  yielding the general
Yukawa interactions
\begin{equation}
{\cal L}_Y = - \sum_{k=1}^2 [ {Y}^d_k{\bar{Q}}'_L\cdot\Phi_k{D}'_R +
{Y}^u_k{\bar{Q}}'_L\cdot\tilde{\Phi}_k{U}'_R ] + {\rm h.c.}, 
\label{Yuk}
\end{equation}
where primes denote the weak basis, $\tilde{\Phi} = i\sigma_2\Phi$, 
and ${Y}^q_k$ denote four 3$\times$3
real Yukawa matrices. Expanding around the ground state (\ref{VEV}),
one obtains the
non-diagonal complex mass matrices 
\begin{equation}
{M}_u = \frac{v_1}{\sqrt 2}{Y}^u_1 + 
\frac{v_2e^{i\vartheta}}{\sqrt 2}{Y}^u_2, \qquad
{M}_d = \frac{v_1}{\sqrt 2}{Y}^d_1 + 
\frac{v_2e^{i\vartheta}}{\sqrt 2}{Y}^d_2 .
\label{MASS}
\end{equation}
It follows that ${M}_u{M}^\dagger_u$ and ${M}_d{M}^\dagger_d$ are arbitrary hermitian matrices.
Diagonalization of these matrices  leads to  charged weak quark interactions of the usual form 
(\ref{CC}) with a complex mixing matrix $V_{KM}$ whose CP phase depends on $\vartheta$.\\
In short, these models have only one CP parameter, namely the 
``vacuum phase angle" $\vartheta$, but a rich
CP phenomenology:\\
i) CPV in charged weak current reactions 
($W^\pm$ and $H^\pm$ exchange) and \\
ii) CPV mediated by flavour-conserving and by flavour-changing neutral Higgs boson $\varphi$ exchange.\\
Note that the observed CPV in $|\Delta S| = 2 
$ $ K^0$-${\bar K^0}$ mixing is dominantly
generated by tree-level $\varphi$ exchange,
$s{\bar d}\leftrightarrow {\bar s}d$. In this respect, these models may be regarded as a 
realization of Wolfenstein's ``superweak hypothesis". 
But in general there  will be also other
CPV  $|\Delta F| = 2 $ tree-level transitions. 
The problematic feature of these models is that fine tuning
of the flavour-changing neutral current couplings (or the appeal to some approximate
symmetry in flavour space)  and/or rather large
$\varphi$ masses are required in order to avoid conflict with experiment 
(leaving aside the strong CP problem).
\par 
The simplest SM extension with spontaneous CPV $and$  
flavour conservation in neutral Higgs particle interactions at
tree-level seems to be the three Higgs-doublet model of 
ref. \cite{Wein3}. CPV originates from  two CPV vacuum phase angles $\vartheta_1,\vartheta_2$ 
which lead to
CPV neutral Higgs mixing and to a complex mass matrix for the charged Higgs bosons.
However, in this model $V_{KM}$ remains real \cite{Bran3} and the
observed CP violation in neutral kaon mixing must be accounted for
by charged Higgs boson exchange (one-loop box diagrams). 
The model appears to be incompatible with experimental
data: One has a hard time explaining why CPV charged Higgs boson
exchange generates $\epsilon \approx 10^{-3}$ in the $K$ meson system,
but suppresses $\epsilon'/\epsilon$ to the level of $10^{-3}$ or even below that number
and the neutron electric dipole moment below  $10^{-25}e$cm.
\par
Is spontaneous CPV a viable concept for supersymmetric extensions of 
the SM? Let us first consider the MSSM and assume CP invariance
of the Lagrangian ${\cal L}_{MSSM}$. It is clear from the discussion 
below eq. (\ref{VLee}) that there is no spontaneous CPV at tree level,
because the couplings $\lambda_1 = \lambda_2 = \lambda_3 = 0$ in the
tree level potential (\ref{Vsusy}). Chargino, neutralino,
$t$, and $\tilde{t}$ contributions to the effective potential
at one-loop \cite{PomHaba} generate small, real couplings
\begin{equation}
\lambda_{1,2,3} \sim g^4/32\pi^2 \sim 5\times 10^{-4} . 
\label{Bed}
\end{equation}
If the parameters are such that condition (\ref{cond})
is fulfilled then the model can have a stable ground state \cite{PomHaba}
which 
spontaneously breaks CP. It follows from eqs. (\ref{cond}) and
(\ref{Bed}) that this requires the parameter $\kappa$ to be small,
$\kappa = {\cal O}(\lambda_1 v^2)$. This implies, however, that the 
lightest of the three neutral Higgs bosons  has
a mass of about $m \approx \sqrt{(4 \lambda_1 v^2 sin^2\vartheta)} \approx$
5 GeV, which is incompatible with experimental
constraints.  Hence this scenario is of no use for the MSSM.
(The appearance of a light boson can be understood from the
Georgi-Pais theorem \cite{GeoP}.)
\par Spontaneous CPV in the next-to-minimal supersymmetric extension of the SM
(see above) was investigated in \cite{Poma,Haba}. Radiative
corrections to the tree-level scalar potential (\ref{VNSSM}), with all couplings now being real 
because of CP invariance, are also essential in this model for the mechanism of spontaneous CPV 
to work \cite{Haba}. (The parameters of the potential are constrained by the requirement that
the masses of the scalar states must be positive.) Refs. \cite{Haba}
find that in this case the mass of the lightest neutral
Higgs boson has an upper bound of about 36  GeV and  the sum of the masses of 
the two lightest neutral Higgs bosons is not much above the mass of the $Z$ boson. It is questionable
whether  this scenario is still compatible with data from LEP. \\ \\ \\
\begin{tabular}{l |c | c}
model        &         mechanism of CPV   &   required non-degeneracy in mass \\ \hline
SM           &         KM           &         u-type quarks, d-type quarks \\ \hline
massive neutrinos   &   KM-like    &           charged leptons, neutrinos \\ \hline

multi-Higgs models  &    neutral Higgs     &    neutral $\varphi_j$ bosons \\
                     &   mixing            &                               \\

                      &  charged Higgs    &       charged Higgs bosons $H^\pm_j$ \\
                       &   mixing         &                                       \\  \hline
MSSM   &  phases in  & scalar fermions of a given flavour \\
    &  SUSY breaking terms & $\tilde{f}_1,\tilde{f}_2$, ($\tilde{f}=\tilde{q},\tilde{\ell}$) \\ \hline
\end{tabular}
\newpage
\subsubsection{Miscellaneous issues}
As has been discussed in section 2.1, CP 
violation \`a la KM requires
non-degeneracy of the masses of both u- and d-type quarks.  
In fact this is a generic feature
of CP violation from the non-gauge sector 
(that is to say, ignoring the $\theta$ term of QCD). For the models discussed
above this is exhibited  in the
 table above. 
Invariants similar to $J_{CP}$ of eq. (\ref{JJ}) can be constructed
 also for the non-KM sources of CPV (cf., for instance, \cite{Inv}).

\par
So far the only hint for CPV beyond 
the KM phase
comes from attempts to develop scenarios for explaining 
the baryon asymmetry of the universe on the basis of
particle physics models and the big-bang model. It is well-known
by now that within this framework baryogenesis requires \cite{Sach}
 baryon
number violation, C and CP violation, and thermal non-equilibrium,
that is to say, an ``arrow of time".
Two types of scenarios have been intensely investigated in recent years:\\
a) Baryogenesis at the electroweak phase transition.
Investigations of the nature of the phase transition lead to the conclusion 
that this scenario does not work within the SM. (For a 
recent review, see \cite{Fodor}.) Even if non-SM interactions 
exist such
that the electroweak phase transition was of first order, 
it is questionable whether KM CP violation did the job. 
(For reviews,
see \cite{Cohen}.) According to present knowledge it seems
that, for instance, two-Higgs doublet extensions \cite{Cohen,Cline} and the 
MSSM \cite{Espin}, with CPV as discussed in section 2, still provide
viable alternatives.\\
b) Out-of-equilibrium decays of ultra-heavy Majorana
neutrinos \cite{FY}, with ($B$-$L$) violation,
 at temperatures far above the electroweak 
phase transition. 
\par It would be fascinating to relate the observed CP violation in the laboratory,
which so far amounts to the $\epsilon$ parameter of neutral $K$ meson mixing,
to the fact that the visible universe contains matter, rather than
anti-matter, with a baryon-to-photon ratio of 
$n_B/n_\gamma \sim 10^{-10}$. However, 
as suggested by the investigations of scenario b), it may be that 
the CP-violating interactions
which were at work in the early universe are irrelevant for 
reactions that can be explored in, say,
 an atomic physics
experiment, at a $B$ meson factory, or even at the LHC. 
In any case, it is challenging enough to search for CPV phenomena
in laboratories on the earth. In the following a
number of those phenomena which are
predicted by the KM mechanism and/or by some non-KM sources of CP 
violation are discussed. 

\section{Weak Decays}
Observable CP violation \`a la KM requires quarks whose weak decays are Cabibbo suppressed. That is not 
the case for $c$ and 
$t$ quarks. Therefore CP searches involving these quarks will predominantly test for new interactions.

\subsection{Kaons and Hyperons}
 
CP violation in 
the kaon system manifests itself in the 
very existence of the decays $K_L \to 2\pi$ and in a non-zero
CP asymmetry, $\delta$,  between the rates of the
 the semi-leptonic decays $K_L\to \pi^\mp\ell^\pm\nu$.  From these observations it can be
inferred that CPV takes place in the 
$|\Delta S|=2$ $K^0$-$\bar{K^0}$ mixing amplitude. The strength of
$|\Delta S|=2$ CPV is characterized by the $\epsilon$ parameter. One has
$\delta \approx 2 {\rm Re}(\epsilon) = $3.27(12)$\times 10^{-3}.$ 
The KM mechanism can naturally explain the order of magnitude of this number -- given the
moduli of the CKM matrix elements measured in other decays. Recall that 
CPV in the SM is small because of small inter-generation mixing angles involved
(cf. eq. (\ref{ImQ})). The parameter $\epsilon$ is determined in the SM by the
ratio of the imaginary part to the real part of the well-known box diagram mixing amplitude.
(To be precise, of its dispersive part). A simple counting
of  the CKM matrix elements involved shows that $\epsilon_{SM} \sim 10^{-3} \sin\,\delta_{KM}$.
\par
The present experimental status of whether
there is also ``direct" $|\Delta S|=1$ CP violation in the $K^0\to 2\pi$ amplitudes is 
inconclusive \cite{eps1,eps2}. New 
experiments \cite{Wahl} aim at measuring the corresponding observable,
namely $\rm{Re} (\epsilon'/\epsilon)$, at the level of $10^{-4}$. On the 
theoretical side
considerable effort has been spent over the last years to calculate the
next-to-leading order QCD corrections to the  effective weak Hamiltonian within the SM, to pursue 
various approaches in 
determining weak matrix elements, and to get a handle on the various uncertainties involved in the 
prediction 
of $ \epsilon'/\epsilon$.
Recent detailed reviews \cite{Buras,Buras1} of the current status estimate 
this  quantity within the SM  $\sim a\ few \times 10^{-4}$. There are considerable 
uncertainties of this estimate due to (a) cancellations of the QCD penguin  
and electroweak penguin diagram contributions to $\epsilon'$, (b) uncertainties
in the knowledge of some SM parameters, notably the mass of the strange quark, and (c) methodic
uncertainties in calculating the non-leptonic weak decay matrix elements.

The present high statistics kaon decay experiments  \cite{Wahl} can also 
search for and investigate several rare $K$ decays. For instance, in the case of
the decay $K_L \to \pi\pi e^+e^-$, 
there is a CP asymmetry in the distribution $d\Gamma/d\phi$ ($\phi$ is the angle between the normal 
vectors of the
$e^+e^-$ and $\pi\pi$ planes) generated by the observed CP violation 
in  $K^0 -\bar{K^0}$ mixing. The asymmetry is predicted to be rather large,
about 14 percent \cite{Sehgal}.

Hyperon decays also offer a possibility to search for CP violation in 
$\Delta S = 1$ decays -- although detectable effects require, very probably,  non SM
CP-violating interactions. Consider for instance the decay of polarized 
$\Lambda\to p\pi^-$ and ${\bar\Lambda}\to {\bar p}\pi^+$.
The differential decay distributions are proportional to 
$(1+\alpha_\Lambda\vec\omega_\Lambda\cdot\hat p_p)$
and $(1+\alpha_{\bar\Lambda}\vec\omega_{\bar\Lambda}\cdot\hat p_{\bar p})$, 
respectively, where $\vec\omega$ 
is the hyperon polarization vector and $\hat p$ is the (anti) proton 
direction of flight in the hyperon rest frame.
The spin analyser quality factor $\alpha$, which is parity-violating, is 
generated by the interference
of S and P wave amplitudes.  CP invariance requires
that $\alpha_\Lambda = - \alpha_{\bar\Lambda}$. Hence a CP observable is 
\begin{equation}
A_\Lambda = \frac{\alpha_\Lambda + \alpha_{\bar\Lambda}}
{\alpha_\Lambda - \alpha_{\bar\Lambda}}.
\label{Aslam}
\end{equation}
Note that $A_\Lambda$ is CP-odd but T-even, i.e., even under the reversal of momenta and spins. Hence 
a non-zero asymmetry (\ref{Aslam}) requires, apart from CP phases, also absorptive parts in the amplitudes.
Neglecting isospin $I=3/2$ contributions, an approximate expression for $A_\Lambda$ 
is given by (see, for instance ref. \cite{Grimus}) 
\begin{equation}
A_\Lambda \simeq - \tan(\delta^P_{1/2}-\delta^S_{1/2})\sin(\varphi^P_{1/2}-\varphi^S_{1/2}),
\label{Aform}
\end{equation}
where $\delta^{S,P}_{1/2}$ and $\varphi^{S,P}_{1/2}$ are the S,P wave final state phase shifts and weak
CP phases for the isospin $I=1/2$ amplitudes, respectively. 

In the Standard Model CP violation in $\Delta S = 1$ hyperon decays is induced by penguin amplitudes. 
Extensions of
the SM may add charged Higgs penguin, gluino penguin contributions, etc.  Predictions for hyperon  
CP observables
like $A_\Lambda$ are usually obtained \cite{Don,HSV,HV} as follows: within a given model of CP violation 
one computes 
first the effective weak $\Delta S = 1$ Hamiltonian at the quark level. 
(In the SM its next-to-leading order QCD 
corrections are known \cite{Buras}.)  The strong phase
shifts $\delta^{S,P}_{1/2}$ are extracted from experimental data. The usual strategy in 
determining the weak 
phases $\varphi^{S,P}_{1/2}$
is to take the real parts of the matrix elements $<\pi p|H_{eff}|\Lambda>^{S,P}_{I=1/2}$ 
from experiment, whereas 
the CPV part is computed using various models for hadronic matrix elements. Although the 
theoretical uncertainties
 are quite large one may conclude \cite{HSV,HV} from these calculations that within the 
SM the asymmetry
 $A_{\Lambda}$ is about $4\times 10^{-5}$.
Contributions from non SM sources of CP violation can yield larger effects, but are  
constrained by the 
$\epsilon'$ and $\epsilon$ parameters
from $K$ decays. He and Valencia conclude that $|A^{non-SM}_{\Lambda}|$ cannot 
exceed $a\ few \times 10^{-4}$.

Data from a high statistics hyperon  experiment \cite{Lambda}  (E871) at Fermilab are presently
being analysed. The decay
chain $\Xi^-\to\Lambda\pi^-\to p\pi^+\pi^-$ and the corresponding decay chain for $\bar\Xi^+$ will be used.
They  $\Xi$'s will be produced  $unpolarized$. Then the $\Lambda$ polarization is given by 
${\vec\omega}_{\Lambda}$=$\alpha_{\Xi}
{ \hat p}_\Lambda$, where ${ \hat p}_\Lambda$ is the $\Lambda$ direction of flight in the $\Xi$ rest frame.
E871 measures the asymmetry 
\begin{equation}
A = \frac{\alpha_\Lambda\alpha_\Xi - \alpha_{\bar\Lambda}\alpha_{\bar\Xi}}
{\alpha_\Lambda\alpha_\Xi + \alpha_{\bar\Lambda}\alpha_{\bar\Xi}} \simeq A_\Lambda + A_\Xi.
\label{AslaX}
\end{equation}
$A_\Xi$ is estimated to be smaller than $A_\Lambda$ because of smaller phase shifts. E871 
expect to produce 
about $10^9$ events. They aim at a sensitivity $\delta A \simeq 10^{-4}$. If an effect will 
be observed at this 
level it will be, in view of the above, most probably of non SM origin. 
 
\subsection{Charm}
$D^0 - \bar D^0$ mixing and associated CP violation in the $\Delta C = 2$ mixing amplitude,
and direct CP violation in the  $\Delta C = 1$ charm decay amplitudes are predicted to be 
very small in the SM. 

In the SM direct CPV may be significant only for singly Cabibbo suppressed decays. 
In this case one has at the 
quark level two contributions to the decay amplitude, namely the usual ``tree" amplitude and the 
penguin amplitude, that  have different
weak phases. At  the hadron level the decay  amplitude is of the form $A e^{i\delta_A} + B e^{i\delta_B}$, 
where $\delta_{A,B}$ are strong interaction phase shifts. This leads to a CP asymmetry

\begin{equation}
A_D = \frac{\Gamma(D\to f) - \Gamma({\bar D}\to{\bar f})}
{\Gamma(D\to f) + \Gamma({\bar D}\to{\bar f})} \propto {\rm Im}(AB^*)\sin (\delta_B - \delta_A).
\label{Asa}
\end{equation} 

Buccella et al. \cite{Buc} have investigated $A_D$ within the SM for a number of Cabibbo suppressed channels.
They calculated the strong phase shifts for the respective channels  by assuming dominance of the nearest 
resonance. For some modes, for instance $D^+\to {\bar K^{*0}}K^+$ and $D^+\to \rho^+\pi^0$ they find
$A_D \sim 10^{-3}$. In some extensions of the SM like non-minimal supersymmetry \cite{Bigi} or 
left-right-symmetric models \cite{LeY} $A_D$ can be larger by about one order of magnitude. 
Moreover,  asymmetries
of the same order could also be generated in these models for Cabibbo allowed and doubly Cabibbo suppressed channels. 

$D^0 - \bar D^0$ mixing  is very small in the SM, $x=\Delta m_D/\Gamma_D << 10^{-2}$. However,
 quite a number of extensions of the SM, for instance multi-Higgs or supersymmetric extensions,   
can lead to $x \sim 10^{-2}$. In these models it is quite natural that there is (new) 
CP violation associated with 
$\Delta C = 2$ mixing. 
It is mostly these expectations \cite{Nir} from   SM extensions that nourish the hope
 of observable mixing and observable indirect and direct CP violation in 
 proposed high statistics charm 
experiments \cite{Kaplan} with $10^8$ to $10^9$ events.

\subsection{Beauty}

High statistics experiments with the aim of measuring CPV rate asymmetries  in 
$B$ decays will provide, in the
years to come,  the decisive tests of the KM mechanism \cite{Buras1,BG}. These asymmetries are characterized by the angles --
conventionally called $\alpha, \beta,$ and $\gamma$ --  of a well-known CKM unitarity triangle,
which visualises the following orthogonality relation of the CKM matrix elements:
\begin{equation}
V_{ud}V^*_{ub} + V_{cd}V^*_{cb}  + V_{td}V^*_{tb} = 0 .
\label{unitar}
\end {equation}
Several fits (see, e.g., \cite{Ali,Pich} and for a more recent 
analysis \cite{Buras1}), using as input the value of $\epsilon$ parameter of the
$K$ system, $B^0_d -{\bar B^0_d}$ 
mixing, etc., have been 
performed to constrain these angles. These fits yield in particular 
$0.3 \le \sin(2\beta) \le 0.8$, 
supporting the expectation that CP violation outside the $K$ system will first be observed through
an asymmetry between the rates of $B^0_d$ and ${\bar B^0_d} \to J/\Psi + K_S$. 
The integrated rate asymmetry, which can be calculated in a clean way (that is to say, 
almost without uncertainties due to hadronic final state interaction phases),
is proportional to $\sin(2\beta).$

Similarly the time integrated rate asymmetry of $B^0_d, {\bar B^0_d} \to \pi^+ \pi^-$ is 
related to $\sin(2\alpha)$.
However, apart from the fact that these modes have very small branching ratios, there is an 
uncertainty in the prediction
of the CP asymmetry because of penguin diagrams contributing to the decay amplitudes. 
In principle this uncertainty 
can be eliminated by an isospin analysis \cite{GL}. (Recall that there is no QCD 
penguin contribution to the $I = 3/2$ component
of the $B_d\to \pi\pi$ amplitude.) The  method requires measuring 
$B^0_d \to \pi^+ \pi^-, \pi^0 \pi^0$ and the
conjugated channels, 
and $B^\pm \to \pi^\pm \pi^0$.  It will be difficult to carry out.

The CP parameter $\sin(2\gamma)$ is for instance related to the time integrated asymmetry of
the rates  $B^0_s, {\bar B^0_s} \to \rho K_S$. However, that is not a clean and feasible way  of
extracting $\sin(2\gamma)$: firstly because these
modes have very small branching ratios and secondly  because of 
theoretical complications in view of penguin contributions. One proposed alternative is as follows \cite{GW}:
From the measured decay rates one has to determine the moduli of the decay amplitudes for
$B^+\to D^0 K^+, {\bar D^0} K^+$, $D_{1,2} K^+$ and for the charge conjugated channels. ($D_{1,2}$ are the
the CP- even and odd eigenstates.) From the two triangle relations relating the three 
complex amplitudes for $B^+$ and for $B^-$, 
respectively,
one can obtain $\sin^2\gamma$ up to an ambiguity which can in principle also be resolved.
(For other proposals to measure the angles $\alpha$ and $\gamma$, see, e.g., the review \cite{Buras1}.) 

According to the  KM mechanism for the three generation SM 
 $\alpha + \beta + \gamma = \pi$. 
A deviation from this relation would provide evidence for new CP-violating interactions \cite{BG2}. 
(If the sum of these angles turns out to be $\pi$, note that this does not necessarily imply absence of
new CPV effects in the $B$ system.)
Of course, more specific searches for new CPV in the $B$ system can 
be made, for instance  by investigating CP observables that
are predicted to be small in the SM, e.g., the  asymmetry  in the rate for 
$B^0_s \to J/\Psi + \phi$ and its conjugated mode and, likewise, the rate
asymmetry for $B^\pm \to J/\Psi + K^\pm$.  (For investigations of non-KM CPV
in $B$ decays, see, e.g., \cite{Worah}.)

\section{Electric Dipole Moments}

The searches for permanent electric dipole moments (EDM), for instance of the neutron
or of an atom with non-degenerate ground state are known to be   a very sensitive  means to trace 
new CPV interactions.
Recall that a non-zero  EDM of a non-degenerate stationary state 
would signal P and T violation, that is,
CP violation assuming  CPT invariance. Consider the expectation value
of the EDM operator ${\bf D} = \int d^3x {\bf x} \rho({\bf x})$, where $\rho$ is
the charge density operator, in  a
particle state $|j>$ of definite total angular momentum at rest.
Rotational invariance tells us that 
\begin{equation}
<j|{\bf D}|j>\, = \, d <j|{\bf J}|j> ,
\label{DJ}
\end{equation}
where ${\bf J}$ is the total angular momentum operator. 
With respect to parity and time-reversal transformations one has
\begin{eqnarray}
{\bf D} \quad \stackrel{P}{\longrightarrow} \quad - {\bf D}, \qquad
{\bf D} \quad \stackrel{T}{\longrightarrow} \quad {\bf D} , \\
{\bf J} \quad \stackrel{P}{\longrightarrow} \quad {\bf J}, \qquad
{\bf J} \quad \stackrel{T}{\longrightarrow} \quad - {\bf J} .
\label{JPT}
\end{eqnarray} 
Hence $d\neq 0$ signals P and T violation.
This argument applies not only to  elementary
particles  but to atoms and molecules as well, as long as the the stationary state under
consideration has no energy degeneracies besides those due to 
rotational invariance. (For an elaborate discussion, see \cite{Rup}.) The experimental signature
for an EDM is a linear Stark effect in an external electric field. 

\par A non-zero atomic EDM $d_A$ could be due to a non-zero electron EDM $d_e$,  non-zero nucleon EDMs, P- and
T-violating nucleon-nucleon, and/or electron-nucleon interactions.
Schematically,

\begin{equation}
d_A  = R_A d_e + C^{eN}_A +C^N_A.
\label{dat}
\end{equation} 
 It has been shown long ago \cite{Sandars} that paramagnetic atoms can have large 
enhancement factors $R_A$. (See also \cite{Commins1} for a recent review.)
More recent atomic physics calculations \cite{Liu} 
 obtained for instance for Thallium the factor $R_{Tl}\simeq -585$ with an estimated error of 
about 10$\%$.
For Thallium one has to good approximation $d_{Tl} \simeq d_e R_{Tl} + C^{eN}_{Tl}$. 
The nuclear contributions
can be neglected for the following reasons: The nuclear 
ground state of $^{205}{\rm Tl}$ has spin 1/2 and therefore cannot have a nuclear 
quadrupole moment. A potential (small) contribution of
a Schiff moment of the Thallium nucleus is irrelevant at the present level of experimental sensitivity. 
From the experimental upper bound \cite{Commins} on $d_{Tl}$ and with $R_{Tl}$ the upper bound
$|d_e| < 4\cdot 10^{-27} e$ cm was derived \cite{Commins}. \\
Very precise experimental upper bounds were obtained on the EDMs of certain diamagnetic atoms, in
particular for mercury \cite{Hg}. The mercury EDM, like that of other diamagnetic
atoms, is not sensitive to $d_e$ but to the Schiff moment
of the $^{199}{\rm Hg}$ nucleus which at the quark-parton level would be due to non-zero (chromo) EDMs of quarks
and/or P- and T-violating quark-quark or gluonic effective  interactions. As the transition from the level
of partons to the level of a nucleus involves large uncertainties the experimental 
limits on the EDMs 
of diamagnetic atoms are difficult to interpret in terms of microscopic models of 
CP violation \cite{Kat}.\\
Experimental searches for a non-zero EDM of the neutron  at Grenoble \cite{Gren} and 
at Gatchina \cite{Len} 
have lead to the upper limit $|d_n|<9\cdot 10^{-26} e$ cm. 

Theoretical predictions of the EDM of the electron -- or of other leptons -- usually constitutes
a straightforward problem of perturbation theory because models of CPV are  weak coupling theories a posteriori.
However, a firm numerical prediction within a given extension of the SM would  require knowledge of
parameters like masses and couplings of new particles, apart from CP phases. 
The calculation of $d_n$ and of T-violating nucleon-nucleon interactions, etc. involves in addition
methodological uncertainties. For a given model of CPV one can usually construct with reasonable precision
the relevant effective P- and T-violating low energy Hamiltonian at the quark gluon level 
which contains
(chromo) EDM operators of quarks, the $G\tilde G$ and $GG\tilde G$ operators, etc. The transition to the nucleon/nuclear level,
that is, the computation of T-violating hadronic matrix elements
involves large uncertainties. In computing/estimating the neutron EDM  naive dimensional estimates,
the  quark and the MIT bag model \cite{McKellar}, sum rule 
techniques \cite{Chem,Khri1,Khri2}, 
and experimental constraints
on the quark contribution to the nucleon spin \cite{Flores} have been used.

As was discussed in section 2.1, the KM phase induces only tiny CP-violating effects in flavour-diagonal 
amplitudes. Hence the SM predicts tiny particle EDMs (barring the strong CP problem of QCD; i.e., assuming
 ${\bar\theta} = 0$).  A typical
estimate \cite{McKellar} for the neutron is $|(d_n)^{KM}| < 10^{-30} e$ cm. In the SM with massless
neutrinos CPV in the lepton sector occurs only as a spill-over from the quark sector: 
estimates \cite{BeSu,Pospel} yield $|(d_e)^{KM}| < 10^{-37} e$ cm.

Quite a number of other CPV interactions are conceivable that lead to neutron and electron  EDMs
of the same order of magnitude as the present experimental upper bounds. 
(For reviews, see \cite{McKellar,BeSu,Barr}.)  Multi-Higgs extensions of the SM can contain
neutral Higgs particles with indefinite CP parity (cf. section 2.3). Exchange of these bosons induces quark and lepton
EDMs already at one loop. For light quarks and leptons the dominant effect occurs at two loops \cite{BaZe}.
In two-Higgs doublet extensions \cite{twol,Hay} of the SM with maximal CPV in the neutral 
Higgs sector and a light
neutral Higgs particle with mass of order 100 GeV neutron and electron EDMs as large as 
$10^{-25} e$ cm and
 $a\ few$ $\times 10^{-27} e$ cm, respectively, can be induced. Contributions from charged Higgs 
boson exchanges can have a similar order of magnitude \cite{Taiwan}.

In the minimal supersymmetric extension of the SM (MSSM) there are in general, apart from the
KM phase, extra CP phases due to complex soft SUSY breaking terms (cf. section 2.3).
 These phases are  not bound to 
be small a priori. They generate quark and lepton EDMs and 
chromo EDMS of quarks at one-loop order \cite{EllFN,Flores,Barb} which can be quite large. (Unless the gaugino,
squark or slepton masses are close \cite{Kiz} to 1 TeV  which causes, however, other problems.) 
In particular, the prediction for the electron, which is not clouded by
hadronic uncertainties, is  $d_e \simeq 10^{-25} \sin{\varphi_e}\ (e$ cm) for neutralino and
$\tilde e$ masses of the order of 100 GeV. That means the leptonic SUSY phase $\varphi_e$ must 
be quite small, $\varphi_e \sim 0.01$,
 which seems unnatural in the generic MSSM case. The constrained  versions 
of the MSSM,
mentioned in section 2.3, lead to substantially smaller predictions for the neutron and electron
EDMs  \cite{Berto,Cott}.
\par In supersymmetric grand unified theories the small phase problem eases by construction, too. 
In the SO(10) model considered in
refs. \cite{Hall,Barb} the phases in the soft terms are assumed to be zero at the Planck
scale. Unification of the quarks and leptons of a generation into a single multiplet leads, apart
from the KM phase, to extra CKM phases entering the fermion-sfermion gaugino (higgsino) 
interactions at the weak scale. 
GIM cancellations lead to a smaller $d_n$ and $d_e$ 
than in the generic MSSM -- but $d_e$ can be close to its experimental upper bound.

Clearly, the present experimental EDM bounds have an impact on the parameter spaces of popular
extensions of the SM. In particular the bound on $d_e$ is important in view of the
``theoretically clean" predictions. 
Further improvement of experimental sensitivity is highly desirable. 
As to future  low-energy T violation experiments:
A number of proposals \cite{Commins1,Hinds,Weis} have been made to improve the experimental sensitivity 
to $d_e$ and to the EDMs of certain
atoms by factors of 10 to 100. The Berkeley Tl experiment \cite{Commins}
will improve its sensitivity to $d_e$ significantly.  An experiment is underway \cite{Hinds}
with the paramagnetic molecule
YbF, which is very challenging but has the incentive of having a high sensitivity to
$d_e$.  There is also a new idea \cite{Golub} to measure the
neutron EDM with substantially improved sensitivity.

The present experimental sensitivity to EDMs of quarks and leptons from the second and third fermion
generation is typically of the order $10^{-16}$ to $10^{-18} e$ cm (see below and \cite{Commins1}).
 Although this is orders of magnitude larger than the present limit on $d_e$ it constitutes 
nevertheless interesting
information. Some CP-violating interactions, for instance CPV Higgs boson or leptoquark exchange, 
lead to EDMs in the heavy flavour sector that are much larger than $d_e$ or $d_n$.

\section{High Energy Searches}

Many proposals and studies for CP symmetry tests in high energetic $e^+e^-$, $p\bar p$,
and $p p$ collisions have been made 
(see  \cite{DV,B1,Gav,Stod,BN} for early
studies). In particular the production and decay of $\tau$ leptons, $b$, and $t$ quarks 
are suitable for this purpose, as it allows for searches of new CPV interactions that become
stronger in the heavy flavour sector. Contributions from the KM phase
to the phenomena discussed below are negligibly small. Typically one pursues statistical tests with 
suitable asymmetries or correlations. Consider a reaction where the 
initial and the final states are eigenstates of CP. This means that the 
various contributions to the scattering amplitude ${\cal T}$, and the observables
associated with this reaction, can be classified as being even or odd under a CP transformation.
CP tests are to be made with CP-odd observables 
$\cal O_{CP}$ which change sign 
under a CP transformation. If the scattering amplitude of the reaction is affected by
CPV interactions in a significant way,  ${\cal T} = {\cal T}_{inv} + {\cal T}_{CPV}$, 
then the interference of the CP-invariant and the CPV part
 generates  a non-zero expectation value
\begin{equation}
 <{\cal O}_{CP}> \quad = \quad \frac{\int d\sigma {\cal O}_{CP}}{\int d\sigma} \quad \neq \quad 0 .
\label{sigma}
\end{equation}
Because an unpolarized $f\bar f$ state is a CP eigenstate 
in its c.m. frame it can be shown \cite{BeNa}
that unpolarized (and transversely polarized) $e^+e^-$ and $p\bar p$ collisions allow for
``theoretically clean" CP symmetry tests: in these cases $<{\cal O}_{CP}>$ cannot be 
faked by CP-invariant
interactions as long as the phase space cuts are CP-blind. 
The ``self conjugate" situation discussed above can  be realized in these cases
by comparing data from the reaction $i\to f$ with those of the
CP-conjugated one $i\to {\bar f}$. In the case of 
$p p$ collisions
potential contributions from CP-invariant interactions  to an observable being used for a CP symmetry 
test (for instance, 
T-odd\footnote{Recall that ``T-odd" refers to being odd 
under the reversal of momenta and spins. The initial and final state 
are not interchanged.} observables will in general receive
contributions from QCD absorptive parts) must be carefully discussed.

In order to maximize the sensitivity to CPV couplings
it is often useful to consider so-called optimal observables \cite{AS} that maximize the signal-to-noise ratio.
For a given reaction and a given model of CPV -- or a model independent description of CPV using
effective Lagrangians or form factors -- with only one or a few small parameters these observables
can be constructed in a straightforward fashion.

\subsection{$e^+e^-\to\tau^+ \tau^-$}

CPV effects in tau lepton production with $e^+e^-$ collisions
 and in $\tau$ decay were discussed
in  \cite{Stod,BN,BBNO,GNelson,Nelson,BGV,AnRi,BBO,Choi,KuM}. 
CPV in 
$e^+e^-\to\tau^+ \tau^-$ can be traced back to non-zero EDM and weak dipole
moment (WDM) form factors \cite{BN,BBNO} $d^{\gamma}_{\tau}(s)$ and 
$d^{Z}_{\tau}(s)$, respectively, where $s = E^2_{c.m.}$. These form factors
induce a number of CP-odd tau polarization asymmetries and spin-spin correlations,
for instance a non-zero $d^{Z}_{\tau}(s)$ (more precisely, the real part of that form factor) 
leads to a difference in the polarizations of $\tau^+$ and $\tau^-$ orthogonal 
to the scattering plane. Because the taus auto-analyse their spins through their parity-violating
weak decays the tau polarization asymmetries and spin-spin correlations transcribe to a number of CP-odd
angular correlations $<{\cal O}_{CP}>$ among the final states from $\tau^+ \tau^-$ decay.

In their pioneering work the OPAL and ALEPH collaborations \cite{O1,O2,A1,A2} at LEP have 
demonstrated that  CP tests in high energy $e^+e^-$ collisions can be performed with an accuracy at 
the few per mill
level. In the meantime the four LEP experiments measured a number of CP-odd correlations in
$e^+e^-\to\tau^+ \tau^-$. They turned out
to be consistent with zero. From these results upper limits on the real and imaginary parts of the 
WDM form factors were derived. The combined upper limit
on the real part is \cite{Wermes} $|{\rm Re}d^{Z}_{\tau}(s=m^2_Z)| < 3.6\cdot10^{-18}e$ cm (95$\%$ CL).

As already mentioned above the tau EDM and WDM form factors can be much larger than the electron EDM.
There are a number of SM extensions where the dominant contributions to these form factors are
one-loop effects, being not suppressed by small fermion masses. In these models
one has $d_\tau  = e\ \delta/m_Z$  with $\delta$ of order $\alpha/\pi$.
For multi Higgs models one finds \cite{BBO} that $d_{\tau}$ can reach $10^{-20}e$ cm, whereas 
CPV scalar leptoquark exchange \cite{BBO} can lead to $d_{\tau}$ as large as $3\cdot 10^{-19}e$ cm.
In \cite{Hollik} the EDM and WDM form factors were computed in the minimal supersymmetric
extension of the SM. These authors obtained $d^Z_{\tau}$
of order $10^{-21}e$ cm.

\subsection{$e^+e^-\to b\ {\bar b}\ gluon(s)$}
CP violation in this neutral current reaction would signal 
new interactions. At the parton level these interactions would affect correlations among
parton momenta/energies and parton spins. While the partonic momentum directions can be 
reconstructed from the  jet directions of flight the spin-polarization of the $b$ quark
cannot, in general, be determined with reliable precision due to fragmentation. This implies
that useful CP observables are primarily those which originate from partonic momentum
correlations \cite{B1}. With these correlations only chirality-conserving effective couplings
can be probed with reasonable sensitivity. Several correlations were proposed and 
studied \cite{B1,KO,B3,AL}. This situation is in contrast to $\tau^+\tau^-$ and $t\bar t$ production
(see below) where the fermion polarizations can be traced in the decays. That is why in these cases
searches for CPV dipole form factors, which are chirality-flipping, can be made with good
precision.

In the framework of ${\rm SU}(2)_L$-invariant effective Lagrangians it can be shown 
that chiral invariant CPV effective
$Zb{\bar b}G$ interactions of dimension $d=6$ (after spontaneous symmetry breaking)
exist \cite{B1,BeNa}. In multi-Higgs extensions of the SM these interactions can be induced to
one-loop order \cite{BBHN}. They remain non-zero in the limit of vanishing $b$ quark mass.
Note that these CPV effective interactions are chiral-invariant $and$ flavour-diagonal which is a remarkable
feature. A dimensionless coupling $\hat h_b$ associated with these interactions \cite{B3}
turns out to be of the order of a typical one-loop radiative correction, i.e., a few percent
if CP phases are maximal. This coupling could be larger in models with excited quarks.

At the $Z$ resonance the above reaction provides an excellent possibility to probe for
this type of interactions. The ALEPH collaboration \cite{Alep} has made a CP 
study with their sample of $Z\to b{\bar b} G$ events. They obtained 
a limit of $|\hat h_b|<0.59$ at 95$\%$ CL.

\subsection{Top Quarks and Higgs Bosons}
Because of their extremely short lifetime top quarks decay on average before 
they can hadronize.
This means that the spin properties of $t$ quarks can be inferred with good 
accuracy from their weak decays. (The SM  predicts that $t\to W\ b$  is the main decay mode.)
Like in the case of the tau lepton a number of $t$ spin-polarization and spin-spin
correlation effects may be used to search for non-SM physics. Because of their
heavy mass,  top quarks -- once they are available in sufficiently large numbers --
 will be a good probe of the electroweak symmetry breaking sector through their
Yukawa couplings. In particular they will be a good probe of Higgs sector CP violation.
Many CP tests involving top quarks have been proposed.
These proposals include $t\bar t$ production in high energy $e^+e^-$ 
collisions \cite{BNOS,BSP,KLY,BO,ArSe,GRA,Rind,Pil1,Wien}
and in $p \bar p$ and $p p$ collisions \cite{BM,ScPe,BeBra2,Sch,Nacht,Atw,GradL,Zhou}
at Tevatron and LHC energies, respectively. 
(As already mentioned,  in the latter case no genuine CP tests in the way 
described above can be made. One must carefully discuss and
compute potential fake effects.) Useful channels for these tests
are the final states from 
semi-leptonic decay of both $t$ and $\bar t$
and those from semi-leptonic   (non-leptonic)  $t ({\bar t})$ decay
 plus the charge conjugated channels.
(The charged lepton from semi-leptonic $t$ decay is known to be the most
efficient $t$ spin analyzer. Non-leptonic $t$ decays, on the other hand,
allow for reconstruction of the top momentum.) Observables ${\cal O}_{CP}$
include triple correlations, energy asymmetries, etc. 
and their  optimized versions. Computations of  $<{\cal O}_{CP}>$
have been made in a model-independent way using effective Lagrangians, 
form factor parameterizations
of the $t$ production and decay vertices, and within several extensions of
the SM, notably two-Higgs doublet and supersymmetric extensions. At the
upgraded Tevatron one can reach an interesting sensitivity
to the chromo EDM form factor of the top of about \cite{BM,Nacht,GradL}
$\delta d^{chromo}_t \simeq 10^{-18} e$ cm. Multi Higgs extensions of the SM
can  induce top  EDM, WDM, and chromo EDM form factors of this order
of magnitude  \cite{BSP,Soxu}. The minimal SUSY extension of the SM leads to
smaller predictions for these form factors \cite{BO,Wien3}. EDM and WDM form factors could be 
searched for most efficiently in $e^+e^-\to t\bar t$ not 
far above threshold \cite{BNOS,BO,Rind}. 
It was shown \cite{BO} that,
within two-Higgs doublet  extensions of the SM,
  neutral Higgs sector CP violation induces effects at the percent level
in this reaction.

A possibility to check for CPV Yukawa couplings of the $t$ quark 
would be associated $t {\bar t}$ Higgs boson production. CP effects
can be large \cite{AS2}, but the cross sections are quite small.

If neutral Higgs boson(s) $\varphi$ 
will be discovered and at least one of them 
can be  produced in reasonably large numbers then the CP properties
of the scalar sector could be determined directly by checking
whether $\varphi$ has $J^{PC}=0^{++},\ 0^{-+}$, or whether it
has undefined CP parity as predicted by multi-Higgs extensions of the SM 
with Higgs sector CPV. A number of suggestions and theoretical
studies in this respect were 
made \cite{BBra1,Nor1,Chang,HMK,Kremer,DK,Seg,CoWi,GG,ABB,Pil}.
(Some of them follow the text book descriptions of how to determine
the CP parity of $\pi^0$.) In the fermion-antifermion decay of a
neutral Higgs particle with undefined CP parity CP violation  occurs at $tree$ $level$
and manifests itself in   spin-spin correlations \cite{BBra1}. One of them is
CP-odd  and 
can be as large as 0.5.
These correlations could be traced in $\varphi\to \tau^+\tau^-$ and,
for heavy $\varphi$, in $\varphi\to t \bar t$; for instance, when $\varphi$ is
produced in high-energetic electron positron collisions. In the case
of LHC production, $p p \to \varphi + X\to t \bar t + X$, interference with 
the non-resonant $t\bar t$ background
diminishes the effect  \cite{BeBra2,BBra1}.

A ``Compton collider" realized by backscattering laser photons
off high energy $e^-$ or $e^+$ beams would be an excellent tool
to study Higgs bosons \cite{Zerwas} by tuning the beams to
resonantly produce $\varphi$. The CP properties of $\varphi$ could
be checked by appropriate asymmetries and correlations \cite{Kremer,GG,ABB}.

\section{Summary}

The gauge theory paradigm, which describes  physics so well up to the highest energy scales
presently attainable, suggests that, if there is physics beyond the Standard Theory, there can be
 a number of different types of 
CP-violating interactions which  manifest themselves in different physical situations.
Hence searches for CP violation effects should be made in as many particle reactions as possible.
Present experimental investigations of $K$ decays and of hyperon decays search for  ``direct"
CP violation in $|\Delta S| = 1$ weak transitions at the level of $10^{-4}$. While 
an effect of this order of magnitude can be induced by the KM phase in $K$ decays, it would point 
towards a 
new source of CPV in the case of hyperon decays. 
However, in order to be eventually able to discriminate better between different models of CPV
improved calculations of hadronic matrix elements both for  $K \to 2\pi$ and for non-leptonic
 hyperon decays  are needed .
The decisive tests of the KM mechanism will hopefully be provided by the $B$ meson factories
in the years to come. The searches for a neutron EDM, atomic EDMs, or other T-violation
effects in atoms or molecules remain a unique low energy window to physics beyond the SM.
Searches of non-SM CP violation can also be made at present and future high energy colliders.
Experiments at LEP have already demonstrated that high-energy CP tests can attain sensitivities
at the sub-percent level.
Specifically, if Higgs sector CPV exists, effects of up to a few percent are possible in the
top quark system. Moreover, when Higgs boson(s) will be discovered and eventually produced in large numbers 
it  is also conceivable to study their CP properties directly. 
\par
While at present CP non-conservation may still be considered, from an agnostic point of view, as a curious 
and small effect of mysterious origin in the neutral kaon system, one can be optimistic that, in view of the 
activities outlined  above, we will have a clearer understanding of the cause of this symmetry violation
in the not too distant future.

\section*{Acknowledgements}
I am indebted to E. Commins, E.A. Hinds, S.K. Lamoreaux, and  K.B. Luk for information about
their present and planned experiments.  Moreover, 
I wish to thank A. Brandenburg and P. Uwer for discussions, and the organizers of the 1998 Schladming Winter School,
W. Plessas and his colleagues, 
for inviting me to this pleasant meeting.

\end{document}